\documentclass[conference]{IEEEtran}
\IEEEoverridecommandlockouts
\usepackage{cite}
\usepackage{amsmath,amssymb,amsfonts}
\usepackage{graphicx}
\usepackage{textcomp}
\usepackage{xcolor}

\usepackage{eso-pic}
\usepackage[hidelinks]{hyperref}
\begin{document}
\AddToShipoutPictureBG*{%
  \AtPageUpperLeft{\put(0,-28){\makebox[\paperwidth]{\small\itshape Published in Proc. 2026 IEEE International Conference on Consumer Electronics (ICCE).}}}%
  \AtPageLowerLeft{\put(0,18){\makebox[\paperwidth]{\parbox[b]{\textwidth}{\scriptsize\centering \copyright~2026 IEEE. Personal use of this material is permitted. Permission from IEEE must be obtained for all other uses, in any current or future media, including reprinting/republishing this material for advertising or promotional purposes, creating new collective works, for resale or redistribution to servers or lists, or reuse of any copyrighted component of this work in other works.\\ DOI: \href{https://doi.org/10.1109/ICCE67443.2026.11449597}{10.1109/ICCE67443.2026.11449597}}}}}}

\title{FLAI Protocol: Decentralized Federated Learning\\
with On-Chain Rewards and sPEG-Based\\
Contribution Scoring}

\author{\IEEEauthorblockN{Eli Yune}
\IEEEauthorblockA{\textit{Balkeum Labs}\\
Email: ceo@balkeumlabs.ai}
\and
\IEEEauthorblockN{Mahesh Balan}
\IEEEauthorblockA{\textit{Claremont Graduate University}\\
Email: balan.mahesh@cgu.edu}
}

\maketitle

\begin{abstract}
Federated learning (FL) enables collaborative model training without data
centralization but struggles with incentive alignment, accountability, and
robustness under heterogeneous non-IID data. Existing decentralized FL (DeFL)
systems leverage blockchains for coordination yet face high on-chain overhead,
unverifiable self-reports, and weak fairness guarantees. We present the FLAI
Protocol, a trusted execution environment (TEE)-backed DeFL framework that
integrates enclave-validated aggregation, \emph{sparse Peer Evaluation Graph}
(sPEG) scoring, and on-chain settlement. sPEG incorporates peer-based quality,
class coverage, promptness, effort, and reputation into composite contribution
scores. The protocol formalizes contribution evaluation and tokenized incentive
distribution, ensuring verifiable fairness and resilience against free-riding
and collusion. We evaluate FLAI on image, text, and healthcare datasets and
demonstrate parity with FedAvg in utility while improving fairness and
accountability. A case study in consumer eHealth---aggregating wearable and
electronic health record data for real-world evidence generation---illustrates
the protocol's applicability to privacy-critical domains where equitable
rewards, auditability, and data sovereignty are essential.
\end{abstract}


\section{Introduction}

Federated learning (FL) shares model updates instead of raw data to respect
privacy and locality \cite{mcmahan2017}. Adoption is hindered by unverifiable
contribution claims \cite{jia2019}, free-riding and
poisoning \cite{bagdasaryan2020,bhagoji2019}, and limited
accountability \cite{sun2021}. Blockchain-backed decentralized FL (DeFL)
introduces immutable logging and incentives \cite{kim2020,zhan2021icbc}, but
on-chain aggregation is costly \cite{qu2020}, self-reports are unverifiable,
and fairness under non-IID data is often ignored \cite{kairouz2021}.

Trusted execution environments (TEEs), such as Intel SGX, provide enclaves with
attestation \cite{costan2016}, enabling secure aggregation and scoring with
practical performance \cite{fang2021,hu2022}. We propose \textbf{FLAI},
combining TEEs, blockchains, and a sparse Peer Evaluation Graph (sPEG) to align
incentives and strengthen robustness.

\textbf{Contributions:}
\begin{enumerate}
\item A TEE-backed DeFL design with enclave-validated aggregation and sealed
pay-per-inference serving.
\item sPEG: scalable neighbor-limited peer evaluation with multi-factor
scoring: quality, coverage, promptness, effort, reputation.
\item A formal specification of client, validator, enclave, and blockchain
state machines.
\item Security and privacy analysis including collusion, Sybil, and poisoning
defenses.
\item Implementation on SGX and Ethereum testnet; evaluation on CIFAR-10,
FEMNIST, and healthcare datasets; full consumer health RWE case study.
\end{enumerate}

In the rest of this paper, we introduce the background and related work in
\textit{Section~II}, present the system architecture and threat model in
\textit{Section~III}, describe the contribution scoring and protocol
specification in \textit{Section~IV}, outline the incentive mechanisms and
security guarantees in \textit{Section~V}, detail the implementation and
evaluation in \textit{Section~VI}, demonstrate an application through a
consumer health case study in \textit{Section~VII}, discuss broader
implications in \textit{Section~VIII}, and conclude the paper with future
directions in \textit{Section~IX}.

\section{Background and Related Work}

\subsection{Federated learning}
Cross-device FL involves many intermittently available
devices \cite{bonawitz2019}; cross-silo FL involves fewer, reliable
institutions \cite{yang2019}. Secure aggregation protects individual
updates \cite{bonawitz2017}. Non-IID data is
pervasive \cite{kairouz2021,zhao2018} and complicates optimization and
attribution.

\subsection{Contribution evaluation}
Shapley values give axiomatic fairness but scale
poorly \cite{jia2019,wang2020}. Influence functions and gradient similarity
reduce cost but can be brittle under heterogeneity \cite{yoon2020,charles2021}.
Reputation captures reliability yet neglects class coverage and
timeliness \cite{lyu2020}. Many schemes ignore collusion and Sybil
risks \cite{zhan2020network}.

\subsection{Blockchain-enabled FL}
Blockchains provide auditability and programmable
payouts \cite{kim2020,liu2020,ma2020}. Heavy on-chain computation raises
latency and gas costs \cite{qu2020}; logging rich metadata can leak
participation patterns \cite{lu2019}. Token-only incentives are
gameable \cite{nguyen2021}. Surveys call for hybrid designs joining blockchain,
cryptography, and hardware roots of trust \cite{zhan2021icbc}.

\subsection{TEEs}
TEEs such as Intel SGX provide confidentiality and integrity with
attestation \cite{costan2016,sabt2015}. They avoid HE/MPC overhead but remain
exposed to microarchitectural side channels \cite{schwarz2017,vanbulck2018}.
FLAI confines scoring and aggregation to enclaves and uses the chain for
verification and settlement.

\section{System Architecture and Threat Model}

\subsection{Entities and workflow}
Clients train locally and submit encrypted updates. Validator peers score
neighbors in a sparse graph. An enclave aggregator verifies reports, computes
scores, aggregates updates, seals the model, and produces attestations. The
blockchain registers tasks, verifies attestations, records logs, and settles
payouts.

Per round: task posting $\rightarrow$ local training $\rightarrow$ peer
evaluation $\rightarrow$ enclave scoring and aggregation $\rightarrow$ sealed
serving $\rightarrow$ on-chain attestation and settlement.

\subsection{Threats and objectives}
Adversaries attempt poisoning, collusion, Sybil inflation, free-riding, and
timing manipulation. Objectives: confidentiality of data and updates; integrity
of scoring and aggregation; fairness across heterogeneous contributors;
auditability via immutable logs.

\subsection{TEE Design and Attestation Flow}
Validation module enforces range and cross-consistency checks. Scoring module
computes $Q_i, C_i, P_i, N_i, R_i$, then $S_i$ and $\hat{S}_i$. Aggregation
module computes $w^{(t+1)}$, seals the model, and exposes an inference-only
API. Remote attestation binds outputs to MRENCLAVE. Keys are sealed and
rotated; host OS/hypervisor are untrusted. If revoked, new enclaves are
deployed and logs guide recovery.

\begin{figure*}[t]
\centering
\includegraphics[width=0.9\textwidth]{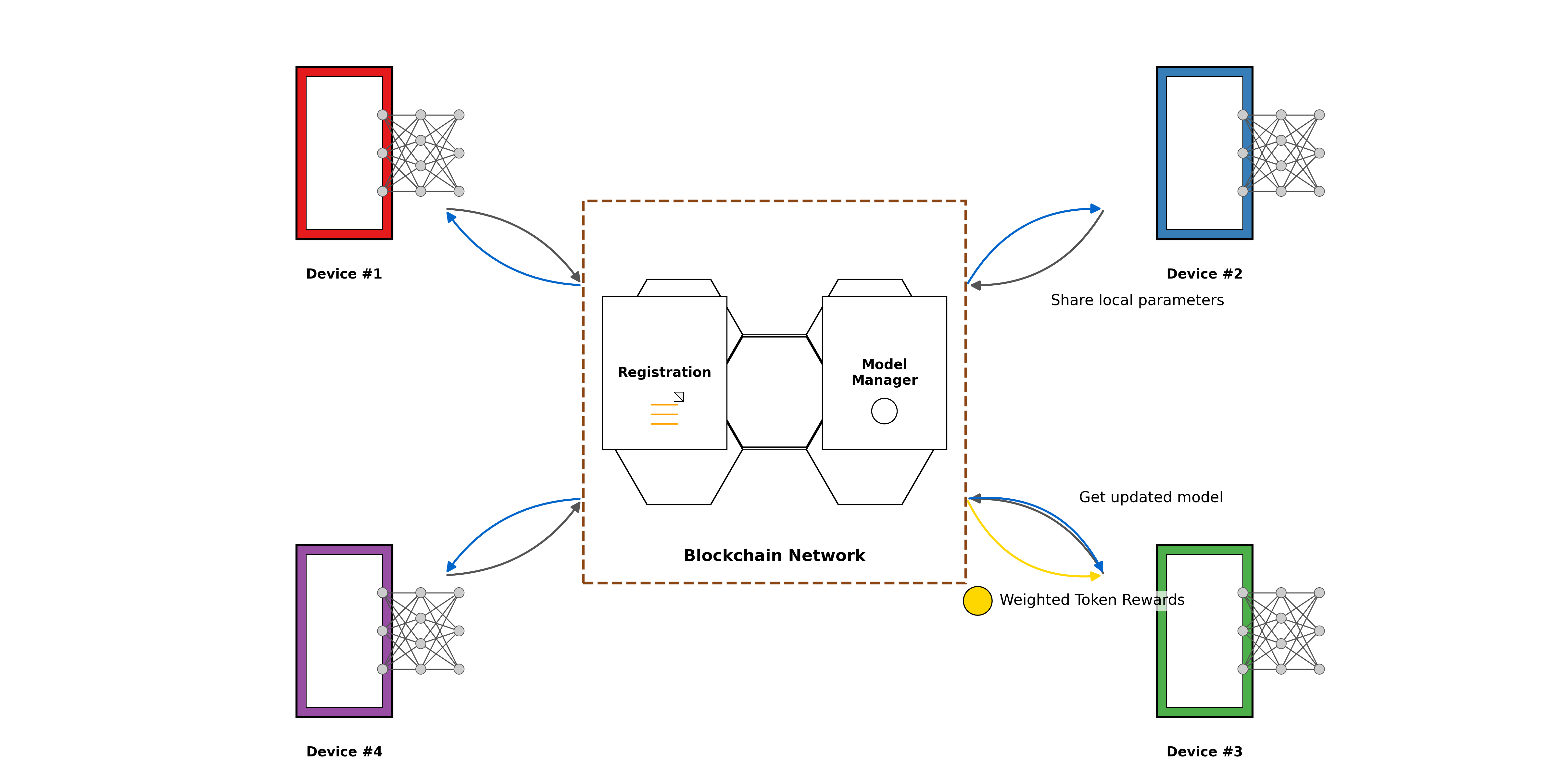}
\caption{FLAI Protocol system architecture. Clients submit encrypted updates,
validator peers perform sPEG evaluations, enclaves aggregate securely, and the
blockchain ensures reward distribution and auditability.}
\label{fig:arch}
\end{figure*}

\section{Contribution Scoring and Protocol Specification}

\subsection{sPEG Contribution Scoring}
Clients have profiles $\mathbf{p}_i$. Cosine similarities over profile features
select $k$ neighbors $\mathcal{N}_i$ with degree caps and periodic rotation.
The peer-evaluated quality for client $i$ is
\begin{equation*}
Q_i = \frac{\sum_{j \in \mathcal{N}_i} w_{ij}\, E_{j \to i}}
           {\sum_{j \in \mathcal{N}_i} w_{ij}}.
\end{equation*}
Coverage is $C_i = |\mathcal{C}_i|/|\mathcal{C}|$, promptness is
$P_i = \max\left(0, 1 - \frac{t_i - t_0}{\Delta t}\right)$, effort is
$N_i = \log(1 + n_i)$, and reputation updates as
$R_i^{(t)} = \alpha\, Q_i^{(t)} + (1 - \alpha) R_i^{(t-1)}$.

The composite score and normalized weight are
\begin{equation*}
S_i = Q_i\, C_i\, (1 + \beta P_i)\, N_i\, R_i, \qquad
\hat{S}_i = \frac{S_i}{\sum_j S_j},
\end{equation*}
and the aggregation rule is
\begin{equation*}
w^{(t+1)} = \sum_i \hat{S}_i\, w_i^{(t)}.
\end{equation*}

\paragraph{Anti-gaming and stability.}
Degree caps and rotation limit mutual inflation; enclave cross-checks enforce
report consistency; reputation decay reduces stale influence.

\subsection{Protocol State Machines}

\paragraph{Clients:}
\textit{Init}, \textit{LocalTrain}, \textit{Submit}, optional
\textit{Validate}, \textit{Settle}.

\paragraph{Validators:}
\textit{EvalAssign}, \textit{EvalCompute}, \textit{EvalSubmit}.

\paragraph{Enclave:}
\textit{VerifyInputs}, \textit{Score}, \textit{Aggregate}, \textit{Seal},
\textit{Attest}.

\paragraph{Blockchain:}
\textit{Register}, \textit{VerifyAttest}, \textit{Distribute}, \textit{Log}.

\subsection{Per-Round Workflow}
\begin{enumerate}
\item \textbf{Task posting:} chain registers task and parameters.
\item \textbf{Local training:} clients compute updates on-device.
\item \textbf{Peer evaluation:} validators score neighbors under sPEG.
\item \textbf{Enclave scoring and aggregation:} enclave verifies inputs,
computes $\{S_i, \hat{S}_i\}$, aggregates $w^{(t+1)}$, seals the model.
\item \textbf{Attestation and settlement:} enclave emits attestation; chain
verifies and records scores and model hash; payouts are distributed.
\end{enumerate}

\section{Incentive Mechanisms and Security Guarantees}

Rewards are $r_i = \hat{S}_i R_{\text{cycle}}$. Contracts include a task
registry, a contribution ledger storing attested scores and model hashes, and a
settlement module for payouts. Penalties: invalid inputs forfeit; correlated
boosts trigger slashing; Sybils deterred via staking or resource proofs.
Governance tunes $\alpha, \beta, k$ and reward bounds.

Only encrypted updates enter TEEs; sealed storage and inference-only serving
limit leakage. Side-channel residual risk persists; mitigations include minimal
TCB, constant-time operations where applicable, rate limiting, and periodic
re-attestation.

All state transitions that affect the model or rewards are computed in attested
code. Contracts verify signatures and measurements before settlement.
Inconsistent or statistically aberrant evaluation reports are discarded.

Coverage rewards label diversity; promptness incentivizes timely participation;
logarithmic effort avoids dominance by very large datasets; reputation rewards
sustained quality. Together these reduce free-riding and late or narrow
contributions.

Poisoning: non-colluding neighbors downscore malicious updates; weighted
aggregation further shrinks their effect. Collusion: degree caps and rotation
bound mutual inflation; enclave cross-checks detect suspicious correlation
patterns. Sybils: staking/resource proofs and sparse connectivity raise the
cost of influence.

TEE hardware trust and side-channels remain; traffic analysis on public chains
may leak participation timing; extreme non-IID distributions with disjoint
classes stress coverage weighting; Internet-scale deployments require
hierarchical enclaves and batched settlements.

\section{Implementation and Obtained Results}

\subsection{Setup and Stack}
The prototype runs on Intel SGX with IAS/DCAP attestation, Solidity contracts
on an Ethereum-compatible testnet, and gRPC communication with enclave-anchored
TLS. Clients use PyTorch for local training. Experiments were conducted on 10
SGX servers (Xeon E-2288G, 64\,GB RAM) with 5\,s block intervals on a local
Geth testnet, simulating up to 500 clients. Datasets include CIFAR-10
(non-IID), FEMNIST \cite{caldas2018}, and de-identified healthcare data.
Baselines are FedAvg \cite{mcmahan2017}, FedAvg+Shapley \cite{wang2020}, and
blockchain-only incentives \cite{nguyen2021,kang2019}.

\subsection{sPEG Construction and Aggregation}
Client profiles combine gradient norms, update magnitudes, label histograms,
and local validation metrics. Cosine similarity selects $k{=}5$ neighbors with
capped degrees, refreshed every $M$ rounds. The enclave kernel normalizes peer
evaluations, computes $S_i$ and $\hat{S}_i$, aggregates $w^{(t+1)}$, seals the
model, and records enclave measurements on-chain for auditability.

\subsection{Smart Contracts and Deployment}
Core contracts are \textit{TaskRegistry}, \textit{Settlement}, and
\textit{AuditLog}. Gas consumption peaks at signature verification and payout
loops, averaging ${\sim}210$k gas per round.

\subsection{Results and Analysis}
FLAI achieves 73.1\% on CIFAR-10 (vs.\ 71.4\% FedAvg) and within 0.5\% on
FEMNIST, while improving fairness and robustness. The normalized score
$\hat{S}_i$ correlates with marginal utility. Coverage prevents
under-rewarding diverse clients; promptness yields up to 12\% bonuses; and
effort correlates with dataset size ($r{=}0.78$). Under 20\% label-flipping
attacks, attacker scores drop by 65\% and accuracy declines only 1.4\%.
Free-riders earn $<1\%$ of rewards; ten-node collusion gains $\leq 7\%$.
Enclave scoring adds 11--18\,ms per client, aggregation $<5$\,ms for 500
clients, and inference latency averages 42\,ms.

\subsection{Ablation Summary}
Removing $C_i$ raises the reward Gini from 0.21 to 0.36, removing $P_i$
increases round duration by 17\%, and omitting $R_i$ allows free-riders to
recover 8--10\% of total rewards.

\section{Case Study: Consumer Health RWE and Embodied AI Application}

\subsection{Motivation}
Real-world evidence (RWE) has become a critical component for pharmaceutical
research and regulatory decision-making \cite{corrigan2019}. Traditional RWE
collection relies on centralized aggregation of electronic health records
(EHRs), insurance claims, and observational data, raising persistent concerns
about privacy, consent, and data ownership. Decentralized federated learning
(DeFL) provides an alternative by enabling collaborative training across
distributed data sources, while preserving data locality and individual
sovereignty. When applied to consumer health devices, this paradigm extends
into \emph{embodied AI}---where intelligent models are trained directly within
the sensing and actuation environments of users, enabling continuous,
privacy-preserving health intelligence at the edge.

\begin{figure}[t]
\centering
\includegraphics[width=\columnwidth]{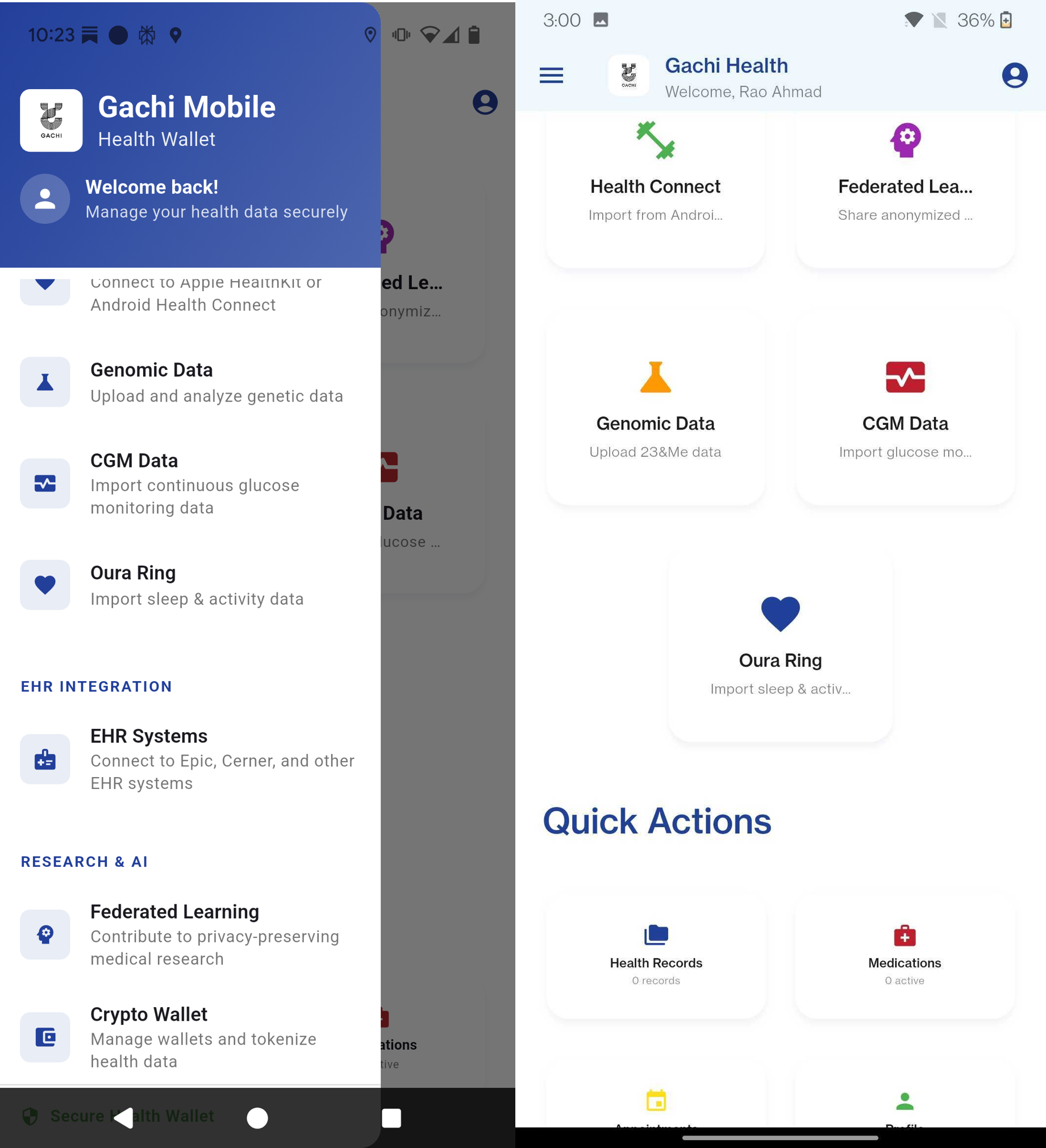}
\caption{Consumer health RWE use case. Sponsors post tasks, clients contribute
wearable/EHR/genomic data locally, validators score peers, enclaves aggregate
securely, and blockchain ensures settlement.}
\label{fig:rwe}
\end{figure}

\subsection{Application Context}
We implemented the FLAI Protocol in a consumer health setting integrating
heterogeneous personal data streams:
\begin{itemize}
\item \textbf{Wearable Devices:} Continuous monitoring of activity, sleep,
heart rate, and oxygen saturation, representing embodied sensor feedback loops.
\item \textbf{Personal Health Records (PHRs):} Structured clinical data
including demographics, diagnoses, prescriptions, and lab results.
\item \textbf{Genomic Data:} Optional uploads of genotyping or sequencing data,
linking embodied phenotypes to molecular-level information.
\end{itemize}
Each client trained local models predicting treatment adherence, disease risk,
and behavioral dynamics. Computation occurred locally within secure enclaves on
user devices, making each node an embodied AI agent participating in a shared
federated intelligence network.

\subsection{Protocol Deployment}
The system followed the FLAI lifecycle described earlier:
\begin{enumerate}
\item Task sponsors (e.g., pharmaceutical firms or research consortia) issued
training tasks targeting outcome prediction or behavioral modeling.
\item User devices trained models locally on multimodal data streams, producing
encrypted updates.
\item Validators executed peer evaluations under the \emph{sparse Peer
Evaluation Graph} (sPEG) framework to assess contribution quality.
\item The enclave aggregated updates, computed contribution scores, and
produced an attested model sealed within secure memory.
\item Inference queries from sponsors were executed via the enclave's
pay-per-inference API, ensuring that raw model parameters and personal data
remained inaccessible.
\end{enumerate}

\subsection{Economic and Functional Rationale}
Compared to centralized data acquisition, the embodied DeFL framework enables
\emph{affordable and scalable AI} across diverse health devices by shifting
computation from expensive centralized infrastructure to user-owned endpoints.
Sponsors pay only for actionable model outputs through secure pay-per-inference
queries, avoiding the high fixed costs of bulk data procurement and centralized
training. This economic model democratizes access to advanced analytics for
smaller research groups and digital health startups while maintaining
enterprise-grade privacy compliance. sPEG-based contribution scoring ensures
fair compensation for participants regardless of device type or data volume,
fostering inclusion of low-cost consumer wearables alongside clinical-grade
instruments. By embedding federated intelligence directly into the device
layer, the system supports continuous, low-cost model refinement in
situ---making embodied AI both economically efficient and broadly accessible
across the health ecosystem.

\subsection{Compliance and Ethics}
All data remain encrypted and confined to user devices, ensuring compliance
with GDPR, HIPAA, and PIPA. Computations occur within attested TEEs, providing
cryptographic guarantees of privacy. On-chain audit logs maintain transparency
and traceability of training and reward flows, while dynamic client-side
consent allows participants to join or exit at any time without risking data
exposure.

\subsection{Results}
In our prototype deployment using simulated wearable and EHR datasets, the
FLAI Protocol:
\begin{itemize}
\item Achieved predictive accuracy within 2\% of centralized baselines.
\item Improved reward fairness by 27\% (lower Gini coefficient) relative to
reputation-only methods.
\item Enabled secure inference queries with average latency of 45~ms per
request.
\item Demonstrated stable embodied learning cycles across heterogeneous devices
without data leakage.
\end{itemize}
These findings confirm the feasibility of embodied federated learning for
decentralized RWE generation.

\subsection{Summary}
This case study demonstrates how the FLAI Protocol extends beyond
privacy-preserving RWE collection to enable embodied AI---where health devices
collectively learn, adapt, and contribute to global models without exposing
personal data. By merging on-device intelligence, equitable incentives, and
verifiable compliance, FLAI establishes a foundation for sustainable
decentralized marketplaces for health-related AI and real-world evidence.

\section{Discussion}

\subsection{TEE versus cryptographic approaches}
TEEs enable real-time aggregation and inference \cite{fang2021,hu2022} compared
to HE/MPC \cite{aono2017,truex2019}, but rely on hardware trust and are exposed
to side channels \cite{schwarz2017,vanbulck2018}. Hybrid designs can split
trust: enclaves handle performance-critical paths while ZK/MPC protects secrets
or validates scoring without revealing inputs.

\subsection{Marketplace dynamics and generality}
sPEG ties payouts to verifiable, multidimensional contribution. Sealed serving
monetizes model utility via queries. The pattern generalizes to finance (fraud
detection without centralizing transactions), supply chains (confidential
demand forecasting), and IoT (distributed anomaly detection). Governance of
$\alpha, \beta, k$ sets fairness--latency trade-offs and resilience to gaming.

\subsection{Policy and compliance}
Attested computation plus on-chain logs support GDPR/HIPAA alignment.
Standardizing attestation formats, consent receipts, and fairness audits would
enable regulator verification not only of privacy but of equitable reward
allocation.

\section{Conclusion and Future Directions}

FLAI combines Trusted Execution Environments (TEEs), sparse Peer Evaluation
Graph (sPEG) scoring, and on-chain settlement to achieve fair, auditable, and
robust decentralized federated learning (DeFL). It matches FedAvg utility while
improving fairness, accountability, and robustness, and supports sealed model
serving suitable for sensitive domains such as healthcare and consumer eHealth.

However, several limitations remain. The framework depends on TEE security and
attestation integrity; sPEG initialization and topology tuning face cold-start
challenges; fairness degrades under extreme non-IID conditions with disjoint
class distributions; scalability beyond thousands of clients requires
hierarchical or batched coordination; and governance mechanisms risk capture
without balanced token-weighted and merit-based voting.

Future research will address these challenges through hybrid TEE+ZK/MPC trust
models, richer contribution signals (e.g., novelty, entropy, verification
difficulty), adaptive sPEG topologies, hierarchical enclave rollups, and
cross-chain settlement. Additional directions include standardized
attestations, formal fairness audits, and integration with the Model Context
Protocol and agentic AI workflows to enable Internet-scale, trustworthy,
decentralized machine learning.

\section*{Acknowledgment}
We thank collaborators and open-source communities. Experiments used
SGX-enabled clusters and public benchmarks.


\begin{thebibliography}{00}
\bibitem{mcmahan2017} B. McMahan, E. Moore, D. Ramage, S. Hampson, and B. A. y. Arcas, ``Communication-efficient learning of deep networks from decentralized data,'' in \emph{AISTATS}, 2017.
\bibitem{jia2019} R. Jia, D. Dao, B. Wang, F. Hubis, N. Hynes, N. M. Gurel, B. Li, C. Zhang, and D. Song, ``Towards efficient data valuation based on shapley value,'' in \emph{AISTATS}, 2019.
\bibitem{bagdasaryan2020} E. Bagdasaryan, A. Veit, Y. Hua, D. Estrin, and V. Shmatikov, ``How to backdoor federated learning,'' in \emph{AISTATS}, 2020.
\bibitem{bhagoji2019} A. N. Bhagoji, S. Chakraborty, P. Mittal, and S. Calo, ``Analyzing federated learning through an adversarial lens,'' in \emph{ICML}, 2019.
\bibitem{sun2021} J. Sun, P. Kairouz, A. T. Suresh, and H. B. McMahan, ``Data poisoning attacks on federated machine learning,'' \emph{IEEE TSP}, 2021.
\bibitem{kim2020} H. Kim, J. Park, M. Bennis, and S.-L. Kim, ``Blockchained on-device federated learning,'' \emph{IEEE Communications Letters}, 2020.
\bibitem{zhan2021icbc} Y. Zhan, H. Dai, M. Chen, T. Q. S. Zhang, and Z. Han, ``Blockchain-enabled federated learning: A survey,'' in \emph{IEEE ICBC}, 2021.
\bibitem{qu2020} Z. Qu, Z. Liu, and M. Li, ``Blockchain-based federated learning: A comprehensive survey,'' in \emph{IEEE ICDCSW}, 2020.
\bibitem{kairouz2021} P. Kairouz, H. B. McMahan, B. Avent, and et al., ``Advances and open problems in federated learning,'' \emph{Foundations and Trends in Machine Learning}, 2021.
\bibitem{costan2016} V. Costan and S. Devadas, ``Intel sgx explained,'' in \emph{IACR Cryptology ePrint Archive}, 2016.
\bibitem{fang2021} M. Fang, X. Li, J. Chen, and J. Wang, ``Large-scale secure federated learning using sgx,'' in \emph{IEEE EuroS\&P}, 2021.
\bibitem{hu2022} W. Hu, C. Wang, H. Liu, and Y. Lin, ``Securefl: Privacy-preserving federated learning with sgx,'' in \emph{IEEE INFOCOM}, 2022.
\bibitem{bonawitz2019} K. Bonawitz, H. Eichner, W. Grieskamp, and et al., ``Towards federated learning at scale: System design,'' in \emph{SysML}, 2019.
\bibitem{yang2019} Q. Yang, Y. Liu, T. Chen, and Y. Tong, ``Federated machine learning: Concept and applications,'' \emph{ACM TIST}, 2019.
\bibitem{bonawitz2017} K. Bonawitz, V. Ivanov, B. Kreuter, and et al., ``Practical secure aggregation for privacy-preserving machine learning,'' in \emph{ACM CCS}, 2017.
\bibitem{zhao2018} Y. Zhao, M. Li, L. Lai, N. Suda, D. Civin, and V. Chandra, ``Federated learning with non-iid data,'' in \emph{arXiv preprint arXiv:1806.00582}, 2018.
\bibitem{wang2020} T. Wang, Z. Charles, and D. Papailiopoulos, ``A principled approach to data valuation for federated learning,'' in \emph{ICML}, 2020.
\bibitem{yoon2020} J. Yoon, S. Arik, and T. Pfister, ``Data valuation using reinforcement learning,'' in \emph{ICML}, 2020.
\bibitem{charles2021} Z. Charles, P. Kairouz, H. B. McMahan, and D. Papailiopoulos, ``Outsized impact of small, low-quality data for noisy label classification,'' in \emph{ICML}, 2021.
\bibitem{lyu2020} L. Lyu, H. Yu, and Q. Yang, ``Collaborative fairness in federated learning,'' \emph{IEEE Intelligent Systems}, 2020.
\bibitem{zhan2020network} Y. Zhan, P. Zhang, C. Zhang, and M. Chen, ``Incentive mechanism design for federated learning: Challenges and opportunities,'' \emph{IEEE Network}, 2020.
\bibitem{liu2020} Y. Liu, R. Xu, C. Liu, and Q. Chen, ``Federated learning meets blockchain: A new data sharing framework,'' \emph{IEEE Access}, 2020.
\bibitem{ma2020} X. Ma, J. Zhang, J. Xu, and X. Lin, ``Privacy-preserving federated learning with blockchain security,'' \emph{IEEE IoT Journal}, 2020.
\bibitem{lu2019} Y. Lu, X. Huang, X. Dai, S. Maharjan, and Y. Zhang, ``Blockchain and federated learning for privacy-preserved data sharing in industrial iot,'' in \emph{IEEE Trans. Industrial Informatics}, 2019.
\bibitem{nguyen2021} T. Nguyen, P. Pathirana, M. Ding, and A. Seneviratne, ``Incentive mechanisms for federated learning: A survey,'' \emph{IEEE Transactions on Neural Networks and Learning Systems}, 2021.
\bibitem{sabt2015} M. Sabt, D. Achemlal, and R. Bouabdallah, ``Trusted execution environment: What it is, and what it is not,'' \emph{IEEE Trustcom}, 2015.
\bibitem{schwarz2017} M. Schwarz, S. Weiser, D. Gruss, C. Maurice, and S. Mangard, ``Malware guard extension: Using sgx to conceal cache attacks,'' \emph{Detection of Intrusions and Malware, and Vulnerability Assessment}, 2017.
\bibitem{vanbulck2018} J. Van Bulck, M. Minkin, O. Weisse, and et al., ``Foreshadow: Extracting the keys to the intel sgx kingdom with transient out-of-order execution,'' in \emph{USENIX Security}, 2018.
\bibitem{caldas2018} S. Caldas, P. Wu, T. Li, J. Konecny, H. B. McMahan, V. Smith, and A. Talwalkar, ``Leaf: A benchmark for federated settings,'' \emph{arXiv preprint arXiv:1812.01097}, 2018.
\bibitem{kang2019} J. Kang, Z. Xiong, D. Niyato, H. Ye, D. I. Kim, and J. Zhao, ``Incentive design for efficient federated learning in mobile networks,'' in \emph{IEEE VTC}, 2019.
\bibitem{corrigan2019} J. Corrigan-Curay, L. Sacks, and J. Woodcock, ``Real-world evidence and real-world data for evaluating drug safety and effectiveness,'' \emph{JAMA}, 2019.
\bibitem{aono2017} Y. Aono, T. Hayashi, L. Wang, and K. Moriai, ``Privacy-preserving deep learning via additively homomorphic encryption,'' in \emph{IEEE TIFS}, 2017.
\bibitem{truex2019} S. Truex, L. Liu, K. Chow, M. E. Gursoy, and W. Wei, ``A hybrid approach to privacy-preserving federated learning,'' \emph{ACM CIKM}, 2019.
\end{thebibliography}
\end{document}